\def\year{2021}\relax
\documentclass[letterpaper]{article} % DO NOT CHANGE THIS
\usepackage{aaai21}  % DO NOT CHANGE THIS
\usepackage{times}  % DO NOT CHANGE THIS
\usepackage{helvet} % DO NOT CHANGE THIS
\usepackage{courier}  % DO NOT CHANGE THIS
\usepackage[hyphens]{url}  % DO NOT CHANGE THIS
\usepackage{graphicx} % DO NOT CHANGE THIS
\usepackage{natbib}  % DO NOT CHANGE THIS AND DO NOT ADD ANY OPTIONS TO IT
\usepackage{caption} % DO NOT CHANGE THIS AND DO NOT ADD ANY OPTIONS TO IT
\usepackage{amsmath}
\usepackage{amssymb}
\usepackage{adjustbox}
\usepackage{subfigure}
\usepackage{cleveref}

\pdfoutput=1

\title{Acoustic Sensing-based Hand Gesture Detection for Wearable Device Interaction}

\author {
    Bing Zhou, Matias Aiskovich, Sinem Guven\\
}
\affiliations {
    IBM Research\\
    \{bing.zhou, matias.aiskovich, sguven\}@ibm.com
}

\begin{document}

\maketitle

%% A teaser figure can be included as follows, but is not recommended since
%% the space is now taken up by a full width abstract.
\begin{figure*}[!ht]
    \centering
     \includegraphics[width=\linewidth]{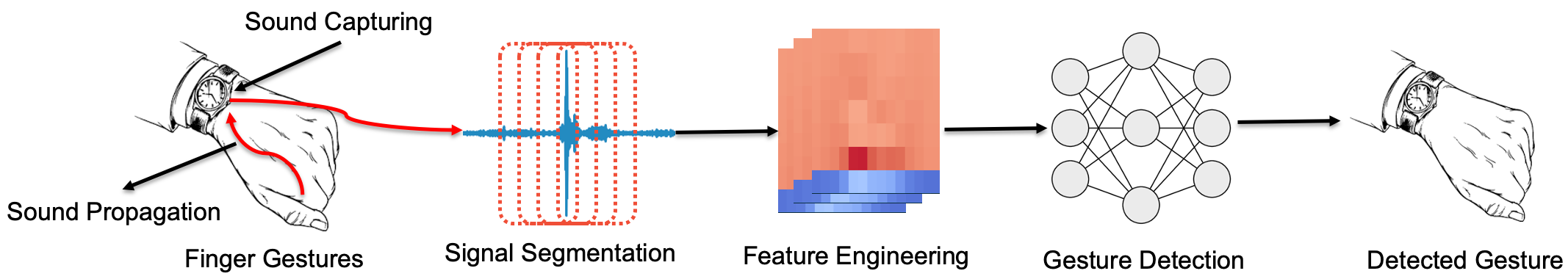}
     \caption{As the fingers move while performing different gestures, smartwatch microphone picks up sounds conducted by bones that emanate from fingers to wrist, and a machine learning pipeline recognizes associated gestures for interaction.
     }
     \label{fig:teaser}
\end{figure*}

%% Abstract section.
\begin{abstract}
%Wearable devices, such as smartwatches and wristbands, have become more popular in recent years. However, they are still difficult to interact with due to small screen real estate, especially when the hand interacting with the device is occupied.
Hand gesture recognition attracts great attention for interaction since it is intuitive and natural to perform. 
%Previous work has explored using radar chips, EMG and pressure sensors for detecting hand gestures, and has demonstrated great potential. However, they all require specialized hardware, which brings additional cost and potential discomfort for users. Built-in inertial sensors on mobile devices have also been explored for gesture detection. However, they can only detect coarse gestures with large hand/arm movements, or detect a small set of hand gestures only if the user's arms are kept static, which heavily constraint their practical use. 
In this paper, we explore a novel method for interaction by using bone-conducted sound generated by finger movements while performing gestures. We design a set of gestures that generate unique sound features, and capture the resulting sound from the wrist using a commodity microphone. Next, we design a sound event detector and a recognition model to classify the gestures. Our system achieves an overall accuracy of 90.13\% in quiet environments and 85.79\% under noisy conditions. This promising technology can be deployed on existing smartwatches as a low power service at no additional cost, and can be used for interaction in augmented and virtual reality applications.
\end{abstract}

%% ACM Computing Classification System (CCS).
%% See <http://www.acm.org/about/class> for details.
%% We recommend the 2012 system <http://www.acm.org/about/class/class/2012>
%% For the 2012 system use the ``\CCScatTwelve'' which command takes four arguments.
%% The 1998 system <http://www.acm.org/about/class/class/2012> is still possible
%% For the 1998 system use the ``\CCScat'' which command takes four arguments.
%% In both cases the last two arguments (1998) or last three (2012) can be empty.

% \CCScatlist{
%   \CCScatTwelve{Human-centered computing}{Visu\-al\-iza\-tion}{Visu\-al\-iza\-tion techniques}{Treemaps};
%   \CCScatTwelve{Human-centered computing}{Visu\-al\-iza\-tion}{Visualization design and evaluation methods}{}
% }

%\CCScatlist{
  %\CCScat{H.5.2}{User Interfaces}{User Interfaces}{Graphical user interfaces (GUI)}{};
  %\CCScat{H.5.m}{Information Interfaces and Presentation}{Miscellaneous}{}{}
%}

%% Copyright space is enabled by default as required by guidelines.
%% It is disabled by the 'review' option or via the following command:
% \nocopyrightspace

%%%%%%%%% BODY TEXT
\section{Introduction}\label{sec:intro}
With recent advances in technology and significant reduction in cost factor, wearable devices like smartwatches and fitness wristbands have become increasingly popular as they provide quick access to key  functionality such as making phone calls, replying to messages, and monitoring our health. However, these devices are still difficult to interact with due to the small size of their screens. Users have to navigate menus through the relatively small screen or via the limited physical buttons, which makes the interaction prone to error and inefficient, particularly because of the wide range of differences in users' finger sizes. Additionally, when one hand is occupied (e.g., while holding a shopping bag), it is hard or sometimes not even possible to interact with wearable devices. Thus, a single-hand-based interaction approach is preferred. 

A series of efforts have been undertaken to address this problem. Researchers have explored various specialized hardware and sensors, such as force sensitive resistors~\cite{dementyev2014wristflex}, EMG sensor~\cite{saponas2010making}, electric field sensor~\cite{wilhelm2015ering}, which have shown very promising results. However, they are not widely adopted in consumer products as they require additional hardware, which is not easy to integrate into wearable devices due to their small size factor. %Additionally, such hardware adds to the cost and potentially increases user discomfort when worn for a long time. 
To eliminate these issues, several researchers propose interaction methods that leverage existing hardware of smartwatches, such as inertial sensors (e.g., accelerometer and gyroscope). %Bernaerts~\textit{et al.}~\cite{bernaerts2014office} develop a system that uses the accelerometer sensor on a smartwatch to detect 3 forearm gestures. 
Shen~\textit{et al.}~\cite{shen2016smartwatch} also use inertial sensors in smartwatches to support more gestures and continuous arm tracking. However, these approaches only support arm-level movement detection, and are not fine-grained enough to detect finger movements. %Furthermore, such large movements are obtrusive and cannot be performed without drawing attention to oneself, and thus, are not suitable for natural interaction.% 
Wen~\textit{et al.}~\cite{wen2016serendipity} propose a solution that recognizes a set of fine-grained hand gestures, however, it only works when the user and their arms are stationary, which highly constrains its practical use. 

Another area of focus for gesture detection has been vision-based approaches for Augmented Reality (AR) and Virtual Reality (VR). LeapMotion ~\cite{weichert2013analysis}, Microsoft Hololens~\cite{hololens} %and Oculus Quest~\cite{oculus} 
as well as several research work~\cite{mueller2018ganerated, sridhar2013interactive} all rely on computer vision techniques to detect hand gestures. However, they suffer from the fundamental limitation of the user's hands having to be in the line of sight for the gestures to be recognized. Furthermore, in most cases, such vision-based techniques rely on depth cameras~\cite{weichert2013analysis,  hololens}, which are subject to poor performance in outdoor environments due to saturation of infrared light in sunny conditions. %Thus, none of the existing solutions is suitable for outdoor use. Other research work and systems eliminate the need for depth cameras by using RGB or monochrome camera based approaches~\cite{mueller2018ganerated, segen1999shadow} %\textcolor{red}{one more RGB based citation}, but such systems have the additional limitation that they do not work in dark environments. Finally, traditional sensing with radio signals have shown promising results on body/arm level gesture tracking~\cite{zhao2018through}, however they require hardware setup and are not fine-grained enough to detect hand gestures.

In this paper, we propose a novel user interaction system, which leverages existing microphones on wearable devices and can be readily deployed on most existing smartwatches or wristbands (as shown in Figure \ref{fig:teaser}). The key insight is that performing a certain hand gesture using fingers creates unique sound signals. Such signals are then conducted by the bones from finger tips to the wrist and can be finally captured by microphones on wrist-worn devices. We propose a series of machine learning pipelines, which recognize the hand gestures from the captured sound signals.
The advantage of our design is that it does not require costly special sensors, such as an EMG sensor, or a force sensitive sensor, that take up space and potentially introduce discomfort when worn for a long duration. As existing wearable devices already use microphones for voice interaction (e.g., ``Siri'' on the Apple watches~\cite{siri}), our approach can be easily integrated without incurring additional energy cost. % or privacy concern of ``always on'' sound recording.
%It detects and recognizes a set of finger gestures by sensing from the bone-conducted sound associated with such gestures. 

%To achieve robust, low-power and easy-to-use gesture-based user interaction using acoustic sensing, we need to address several challenges: 1) sound signals created by finger movements are weak, which makes the signal difficult to capture, especially in noisy environments; 2) sophisticated signal processing, feature extraction and machine learning techniques are needed for reliable and accurate gesture recognition; 3) additionally, we need to make sure the  system consumes low power if this is to be an ``always on'' service.

%This paper proposes a system that addresses the above challenges via 
We develop a simple prototype based on a personal computer and an external microphone setup. The results show that we can reliably capture the hand gestures with just an external microphone attached to the wrist. This also demonstrates the feasibility of interacting with smartwatches using their embedded microphones\footnote{We only use smartwatches for sound capturing from some participants in our studies due to lack of smartwatches for all participants and current challenges with co-location when running the experiments.}. Below we summarize the contributions of this paper:
\begin{itemize}
    \item We propose a novel approach for smartwatch interaction leveraging the bone-conducted sound sensing, which requires no additional hardware and can readily be deployed on existing devices.
    \item We design a series of sound signal pre-processing techniques to suppress the background noise to improve the signal-to-noise ratio (SNR), and extract reliable features for further detection. 
    \item To balance the trade-offs between performance and power consumption, we develop a two-stage machine learning pipeline to minimize the impact on battery life.
    \item We design an acoustic data augmentation scheme for generating ``synthesized'' training samples, which reduces false negatives significantly for noisy environments.
    \item We build a prototype and show that our system can be used to perform both 2D and 3D interaction, and achieves an overall accuracy of 90.13\% in quiet environments and 85.79\% in noisy conditions. 
\end{itemize}

%To the best of our knowledge, this is the first work to leverage bone-conducted sound signals created by finger movements for wearable device interaction, demonstrating robust performance without requiring any additional special sensor.

\section{Related Work}
\textbf{Wearable Device Interaction.} 
WristFlex uses an array of force sensitive resistors (FSRs) worn around the wrist, and the interface can distinguish subtle finger pinch gestures~\cite{dementyev2014wristflex}. SkinWatch~\cite{ogata2015skinwatch} provides gesture input by sensing deformation of the skin under a wearable wrist device, such as a smartwatch or wrist band. It takes the two-dimensional deformation signal from a small and thin sensor attached to the user's skin as input, and matches it to some pre-defined input gestures. 
%\textcolor{red}{A gesture command that is matched by learning data and two-dimensional linear input recognizes the gestures. The sensing part is small, thin, and stable, to accept accurate input via a user's skin.}
eRing~\cite{wilhelm2015ering} detects multiple hand gestures with a single ring through electric field sensing. HandSense~\cite{nguyen2019handsense} explores recognizing dynamic, micro finger movements using capacitive coupling for interacting with a head-mounted device. Electrodes are attached to fingertips of users' gloves and the capacitive coupling among all pairs of electrodes is measured quickly to infer the real-time spatial relationship between fingers. 

\textbf{Acoustic Sensing on Smart Devices.} 
Acoustic signals on smart devices have been explored for many applications in recent years, such as distance measurement~\cite{peng2007beepbeep} and tracking~\cite{zhou2017batmapper}.%, and vital signal monitoring. Distance measurement between devices using acoustic signals is used for interaction between users~\cite{peng2007beepbeep}. Liu \emph{et~al.}~\cite{liu2015snooping} demonstrate that it is possible to detect keystrokes from the sound signals generated from typing. EchoTag~\cite{echotag} recognizes the pre-defined locations by sensing the environment with actively emitted sound signals. Zhou \emph{et al.} developed a mobile application~\cite{zhou2017batmapper} to sense the environment shape and track the movement of mobile devices using sound echo signals and inertial sensors.
%Besides, acoustic ranging can significantly improve smartphone localization accuracy, e.g., adding constraints among peer phones~\cite{liu2012push}, deploying an anchor network that transmits spatial beacon signals~\cite{liu2013guoguo}, or enabling high-precision infrastructure-free mobile device tracking~\cite{zhou2017battracker}.
Acoustic sensing is also explored for active finger movement tracking~\cite{nandakumar2016fingerio,wang2016device}. FingerIO~\cite{nandakumar2016fingerio} and LLAP~\cite{wang2016device} actively emit high frequency sound signals from the built-in speakers on mobile devices, and leverage phase shift in received signals for finger movement tracking. However, such approaches usually only support tracking a single finger movement close to the device. They do not recognize hand gestures involving multiple fingers, which limits their applicability in interaction. %Another application of acoustic sensing is  monitoring human respiration due to its high sensitivity to micro movements. For example, ApneaApp~\cite{nandakumar2015contactless} monitors the minute chest and abdomen breathing movements using signal phase shifts. 
Compared to such previous work, our approach leverages bone-conducted sound signals from finger movements for wearable device interaction.
\section{Method}\label{sec:acoustic}
\subsection{Gesture Design}
There are several considerations when designing gestures from finger movements. First, the gestures should generate sound signals that should be distinguishable so that the associated gestures can be reliably detected. Second, these gestures should be subtle enough so that they can be performed without drawing too much attention to the user. Third, the gestures should be intuitive and natural to perform. 

\begin{figure}
  \centering
  \includegraphics[width=3.3in]{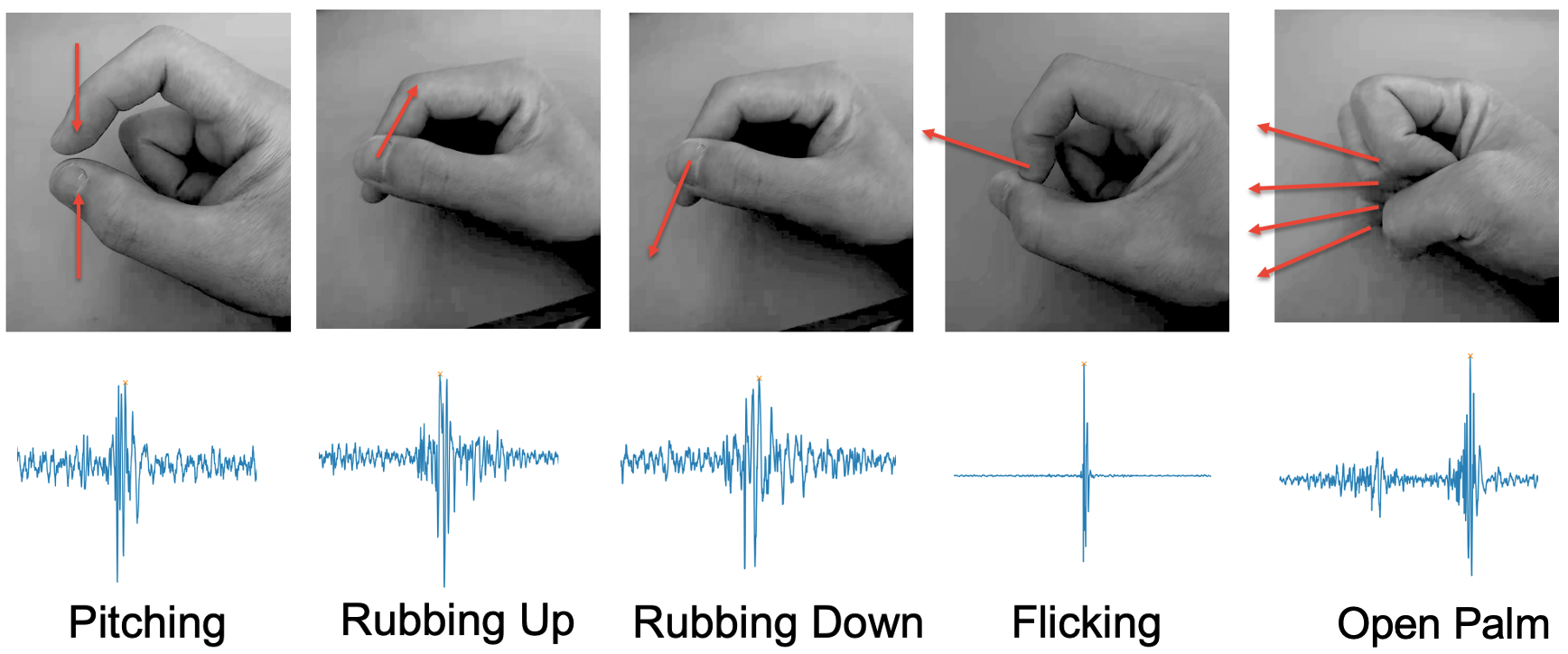}\\
  \caption{Hand gesture set designed for interaction, which includes pitching, rubbing up, rubbing down, flicking and opening the palm, with respective sample sound signals.}\label{fig:gestures}
\end{figure}
In this work, we propose five gestures: pinching, flicking, rubbing up, rubbing down and opening up, as shown in Figure \ref{fig:gestures}. They can be mapped to common interactions on a smartphone or smartwatch, such as click, go back to previous level, scroll up/down, and go back to home, respectively. In Figure \ref{fig:gestures}, we also show a typical sample of the sound signals generated by each gesture's associated finger movements. 
%For pitching, rubbing up/down, the SNR is lower than flicking and open palm due to the lower sound amplitude caused by slighter finger movements. Though the signals from pitching and rubbing up/down look similar in these samples in time domain, they are different in nature. Clicking generates a shorter pulse compared to rubbing fingers, and rubbing fingers up and down creates opposite frequency changing trend in frequency domain. Flicking generates a very sharp pulse with a large amplitude due to the nature of the gesture, which is unique compared to other gestures. Open palm gesture requires two sub-actions, holding the fist closed and then opening it. Thus the signal contains two sub-sequential pulses, one with smaller amplitude due to the holding process and the other one caused be opening the palm, which is shown clearly in Figure \ref{fig:gestures}. 
The complete set of these gestures provides the basic operations necessary to interact with smart devices, such as smartwatches or wristbands.

\subsection{Gesture Detection Architecture}
\begin{figure}
    \centering
    \includegraphics[width=3.2in]{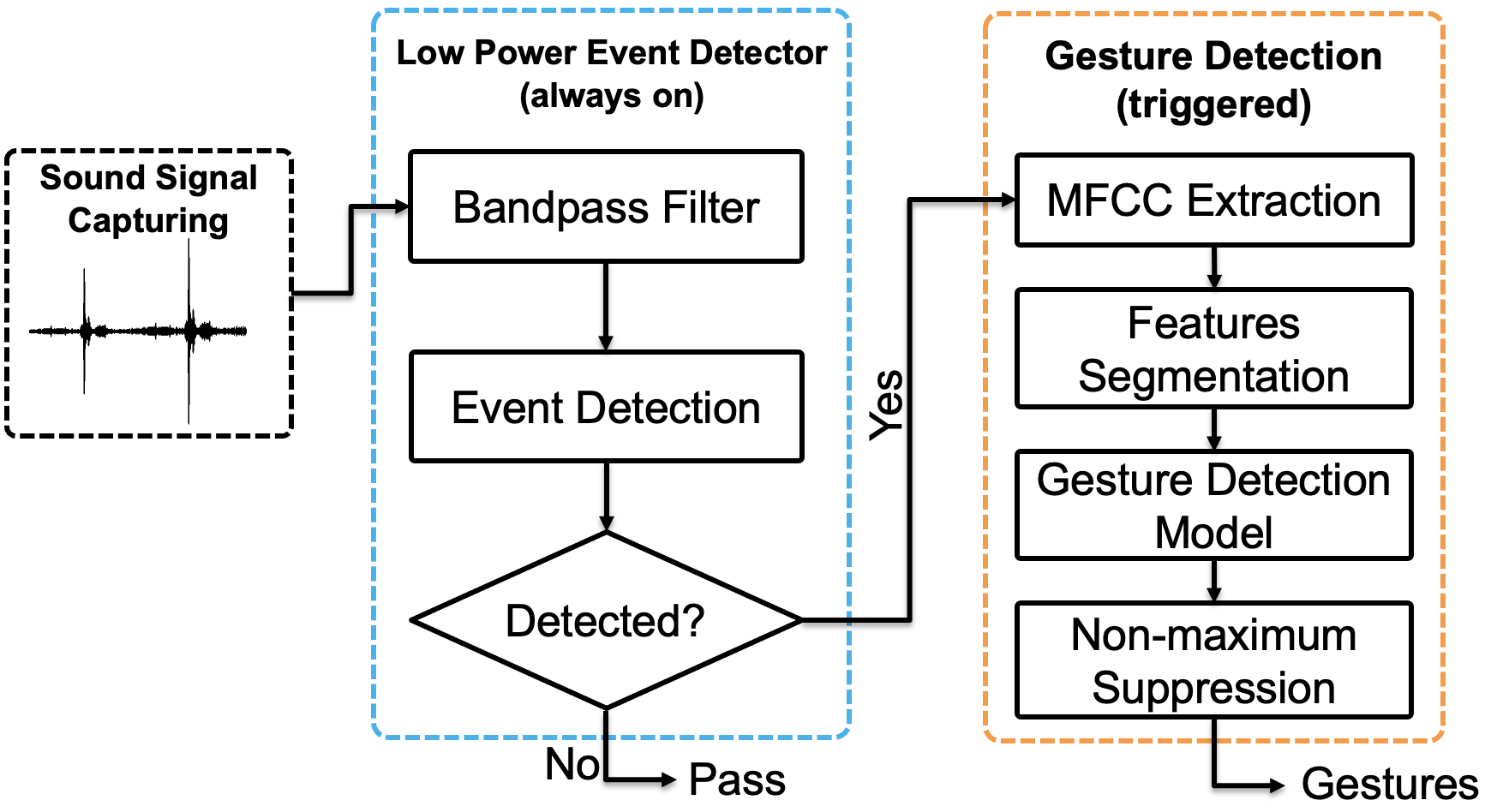}
    \caption{Our design consists of three major components: a signal capturing module, a low power always on module for signal filtering and event detection, and a triggered module for more computational heavy feature extraction and gesture model inference.}
    \label{fig:method}
\end{figure}

To detect the hand gestures in real-time, we propose a two-stage detection architecture, as shown in Figure \ref{fig:method}, which can be divided into three modules: \textit{sound signal capturing}, \textit{low power always-on event detection}, and a \textit{triggered module for gesture detection}.

\subsubsection{Sound Signal Capturing}
There are usually one or more microphones on smartwatches for sound recording, speech recognition and making phone calls. For example, Apple watch has a microphone enabled as ``always on'' for its Siri function. Since the smartwatches are attached to the wrist, the sound can be captured via bone-conduction propagated from fingers to the wrist, which provides better signal-to-noise ratio than air-conducted sound. By accessing the microphone, we get a stream of real-time sound signals, which are fed to our pipeline for gesture detection.

\subsubsection{Low Power Event Detector}
To avoid continuous computation of heavy feature extraction and deep model inferences, we design a low power event detector to detect possible finger gestures, and trigger the gesture classification only when an event is detected. This step avoids unnecessary computations when there is no hand gesture happening. We apply a few lightweight steps to do the first-stage event detection, which consists of two steps: signal filtering and signal detection. This module keeps running in the background so that every intended finger gesture can be captured. We have intentionally designed this ``always on" approach compared to, say ``Siri" like voice command-based activation of gesture recognition, as we believe that interaction should be subtle, natural and applicable in any environment including circumstances where voice-based commands may not be appropriate (e.g., while watching a movie at a movie theater). 

\begin{figure}
    \centering
    \includegraphics[width=2.5in]{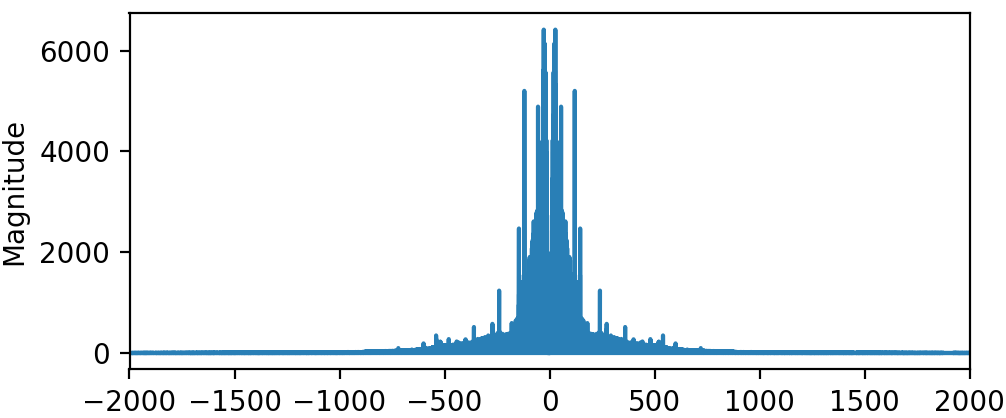}
    \caption{Frequency analysis of sound signals of hand gestures.}
    \label{fig:freq}
\end{figure}

\textbf{Signal Filtering.}
%The microphone in the smartwatches captures the sound propagated from fingers into a stream of instantaneous waveform samples, at a sampling rate of 44100 $Hz$.
Before we further process the raw sound signal, we need to filter out potential background noise to improve the SNR. We perform such filtering by using a bandpass filter, which allows the sound frequency range of the gestures to pass and filters out the background noise. To find out the frequency range of hand gesture sounds, we collect a series of gesture sound signals in a relatively quiet environment to analyze the frequency distribution. The result is shown in Figure \ref{fig:freq}. We can see the frequencies mainly fall in the range of 10 - 500$Hz$.
Thus we design a Butterworth band-pass filter~\cite{hussin2016design} with cutoff frequency range 10 - 500$Hz$ to remove the higher frequency background noises so that weak gesture sounds from fingers will not be buried in the background noise.

\textbf{Event Detection.} The signal detection component detects whether there is a possible hand gesture event. This can avoid most of the unnecessary computations when the environment is relatively quiet, or only contains higher frequency sound, which is filtered out by the bandpass filter. Note that the main goal of this first-stage detection is not to guarantee high accuracy, but to rule out obvious non-event periods. To accomplish this, we designed a simple heuristic: we create a buffer with a time window of 2 seconds with a moving step of 0.1 seconds. We detect the peaks above a certain threshold, if a peak is detected in the central portion of the segment (i.e, between 0.5 -- 1 seconds), the signal is locked. We crop the center 1 second window signal and feed it into next stages, which triggers the sequential steps for gesture detection. In our implementation, we chose the threshold empirically based on our experiments to achieve the best balance of false alarms and missed gestures.

\subsubsection{Gesture Detection}\label{sec:training}
Once an event is captured, it triggers the gesture detection module, which consists of three steps: MFCC feature extraction, gesture detection model and non-maximum suppression.

\begin{figure}
    \centering
    \includegraphics[width=2.5in]{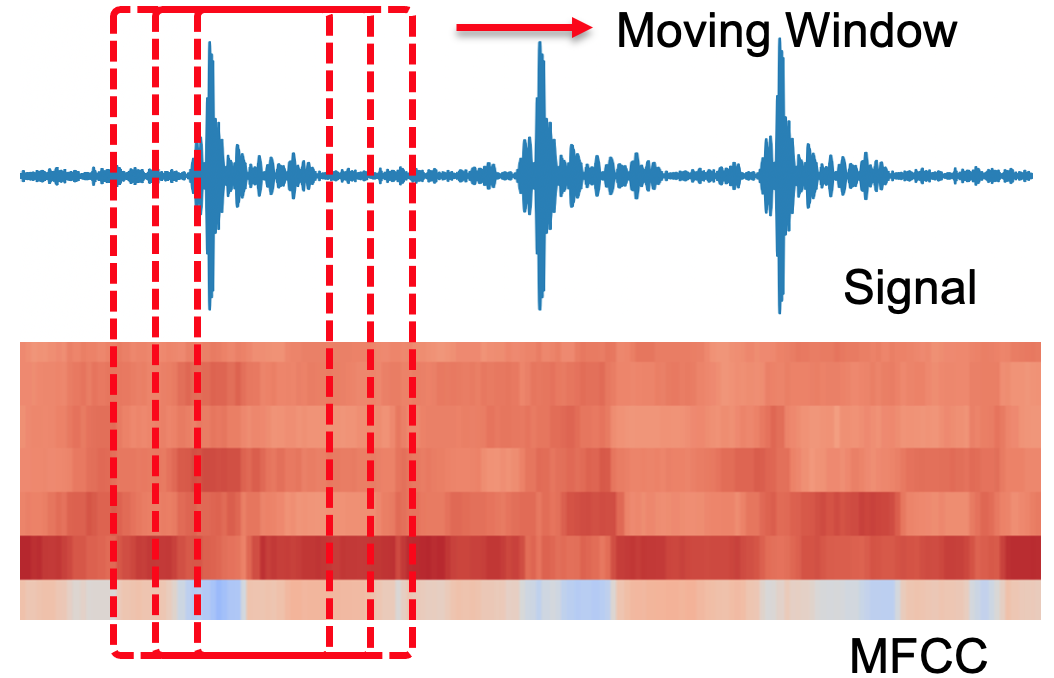}
    \caption{We extract MFCC features from the filtered sound signal and apply a moving window to segment the MFCC features, which are fed into machine learning models for recognition.}
    \label{fig:window}
\end{figure}

\textbf{MFCC Feature Extraction.}
After we get the filtered signal, we apply spectrum analysis on the signal to turn it into frequency domain, which provides richer information than the time domain. Specifically, we extract the MFCC features~\cite{logan2000mel} as shown in Figure \ref{fig:window}. This gives us a stream of 2D heat maps (i.e., spectrograms), describing the sound features in frequency domain. Then, we segment the signal with a sliding windows with a step of 0.05 seconds and a window length of 0.5 seconds, which is roughly the duration of a single gesture. The prepared data is then fed into our machine learning pipeline for prediction. For each 0.5 seconds segmented MFCC feature window, our model predicts the probability of each hand gesture.

%\textcolor{red}{PLEASE REPHRASE, esp the PLUS part does not read well}For each feature window (0.5 seconds period), our acoustic model makes a probability distribution over the set of hand gesture classes, plus no-event/background noise.

\begin{figure}
    \centering
    \includegraphics[width=3.0in]{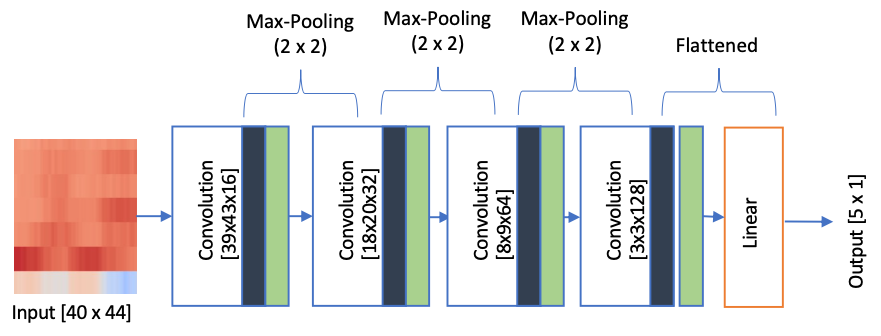}
    \caption{Gesture detection model architecture.}
    \label{fig:net}
\end{figure}

\textbf{Gesture Detection Model.}
%We design an end-to-end machine learning framework for gesture detection, which consists of two major components (shown in \autoref{fig:method}): a CNN based acoustic model for individual prediction and a HMM score aggregation module.
%Traditional acoustic features such as mel-frequency cepstral coefficients~\cite{logan2000mel}, chromagram~\cite{muller2005audio} and spectral contrast~\cite{jiang2002music} have been proven to be effective in human speech recognition and voice-based authentication. 
Recently, deep learning approaches (especially CNNs) have shown a great successes in a variety of challenging tasks, such as image classification, due to their powerful automatic feature extraction~\cite{he2016deep,simonyan2014very}. We design a CNN-based neural network, which takes as input a MFCC of the segmented signals, and trains it on a large data set collected from our test users. Considering the targeted deployment on wearable devices with limited computation resources, we choose to design a light-weight CNN model for detection.
The customized CNN architecture designed for gesture recognition is shown in Figure \ref{fig:net}. The input layer takes the 40x44x1 MFCC features as input, and output the probability of each gesture. Then, we add four consecutive CNN layers for feature extraction. We use rectified linear unit (ReLU) as the activation function for convolutional layers, and after each activation we add a max pooling layer and a dropout layer as regularization factors, to avoid over-fitting.  Finally, we flatten the CNN feature maps followed by two linear layers. The output is a dense layer with a Softmax activation function, which predicts the probability of each gesture class. The CNN is trained on a data set that contains acoustic samples from 5 classes (5 gestures) and categorical cross-entropy is used as the loss function. 

We carefully choose the number of hidden layers and hidden units of each layer of the network to fit the limited computational resources available in wearable devices. The output of the model provides a distribution of scores over each finger gesture for each frame. Due to the sliding window we used, there will be overlaps between frames, so we need to further process the results to generate the final output.

\textbf{Non-maximum Suppression.}
The softmax layer outputs the probability for each gesture as $p_i$ and $\sum_{i=1}^{5}p_i=1$. We output the detected gesture if the probability is larger than a pre-defined threshold $\epsilon$. Since we are using a moving window for detection, there could be multiple results predicted for one actual event.%, as shown in Figure \ref{fig:nonmax}.
To generate one final recognition result for each actual event, we apply non-maximum suppression~\cite{neubeck2006efficient} algorithm to consolidate the results. %Non-maximum suppression algorithm is usually used in object detection tasks in computer vision to generate non-overlapping bounding boxes. We take the same idea and apply it in our problem to generate non-overlapping detection segments. Suppose we get a series of recognition results $\mathcal{P}=\{P_1, P_2, \dots, P_N\}$, where each $P_i=(c_i, p_i, [start_i, end_i])$ and $c_i$ is the gesture label, $p_i$ is the probability, and $[start_i, end_i]$ are the start and end timestamps. The algorithm can be illustrated as follows:

\subsection{Data Collection and Augmentation} 
\textbf{Training Data Collection.} Training a neural network usually requires a large amount of labelled training data, which necessitates a lot of human effort. Additionally, in order for the model to generalize well, the training data should represent different environments. Labelling the subtle sound signals for our application presents a very challenging problem. We propose a method to collect training data in a quiet environment, thus, it can be segmented and labelled automatically. During labelled data generation step, the user repeats the same gesture for each round, and switches to another gesture in the next round. Then the signals are automatically segmented using simple peak detection under the constraint of a minimum interval of 0.5 second in between. Through this step, we can easily get labelled clean data (i.e., with minimum to no background noise). However, training a model on such a dataset may not work well in a more typical environment, which is usually noisy. To solve this problem, we propose an acoustic data augmentation technique to generate more synthesized training data with background noise.

\begin{figure}
  \centering
  \includegraphics[width=3.3in]{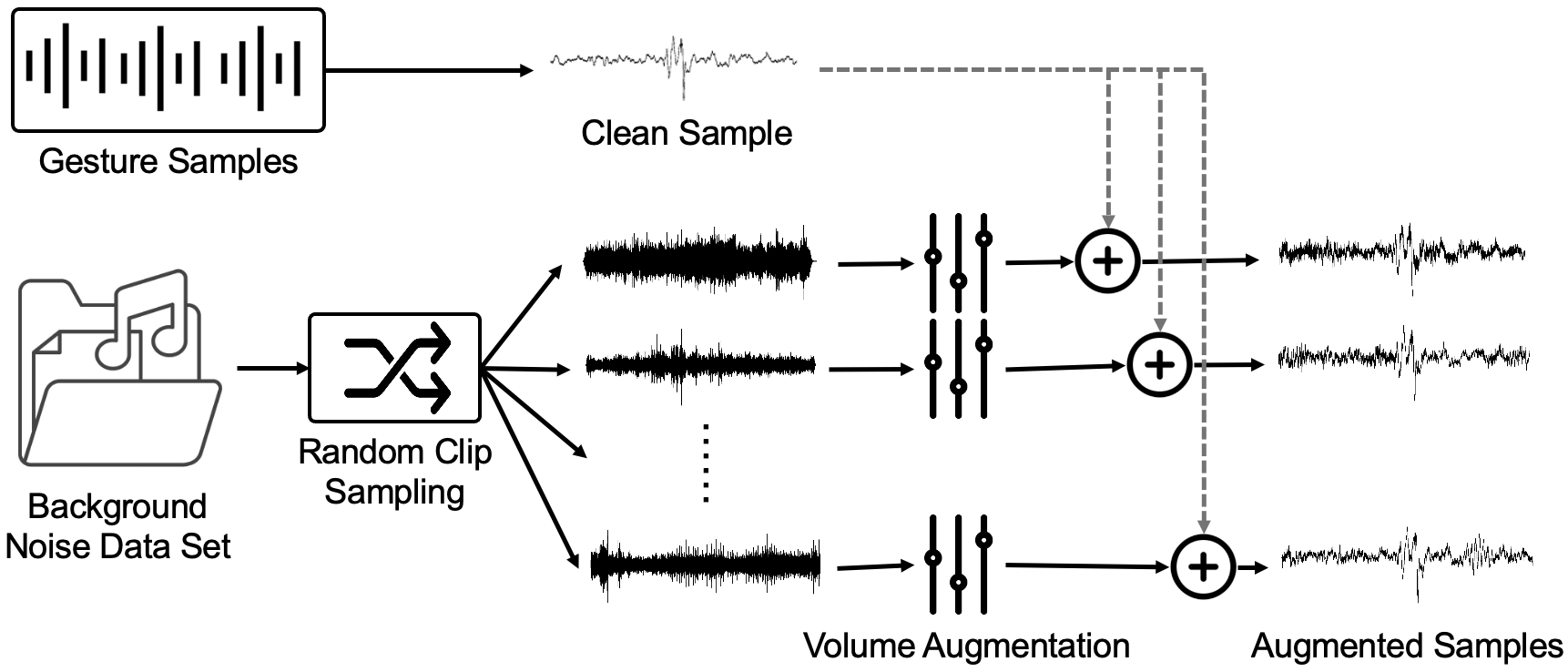}\\
  \caption{Acoustic data augmentation. Multiple augmented samples are created by combining each clean sample with several randomly sampled background noise samples.}\label{fig:augmentation}
\end{figure}

\textbf{Acoustic Data Augmentation.}
Data augmentation is commonly used to increase the amount of data by adding slightly modified copies of already existing data or newly created synthetic data from existing data~\cite{shorten2019survey}. It is an effective way to prevent over-fitting when the amount of data is relatively small. In our case, we have a highly imbalanced data set, i.e., we can have unlimited background noise and our clean training samples are quite limited. To boost the training data set and increase the robustness against background noise, we design an acoustic data augmentation pipeline, shown in Figure \ref{fig:augmentation}. We download a large amount of audio tracks of daily environment conditions as background noise from online sources~\cite{zapsplat} and use them as our background noise data set. Then, for each clean sample of a certain gesture, we augment it to generate multiple augmented samples by adding it with some randomly sampled clips from background noise data set. To better simulate the actual data in different environments, we also augment the background noise volume level by randomly adjusting the volume within a range. With this acoustic data augmentation, we can obtain a large training data set, which has more variety for robust model training and avoids over-fitting. Our evaluation results shows a significant performance increase with augmentation.

\section{Implementation}
\begin{figure}
    \centering
    \includegraphics[width=3.3in]{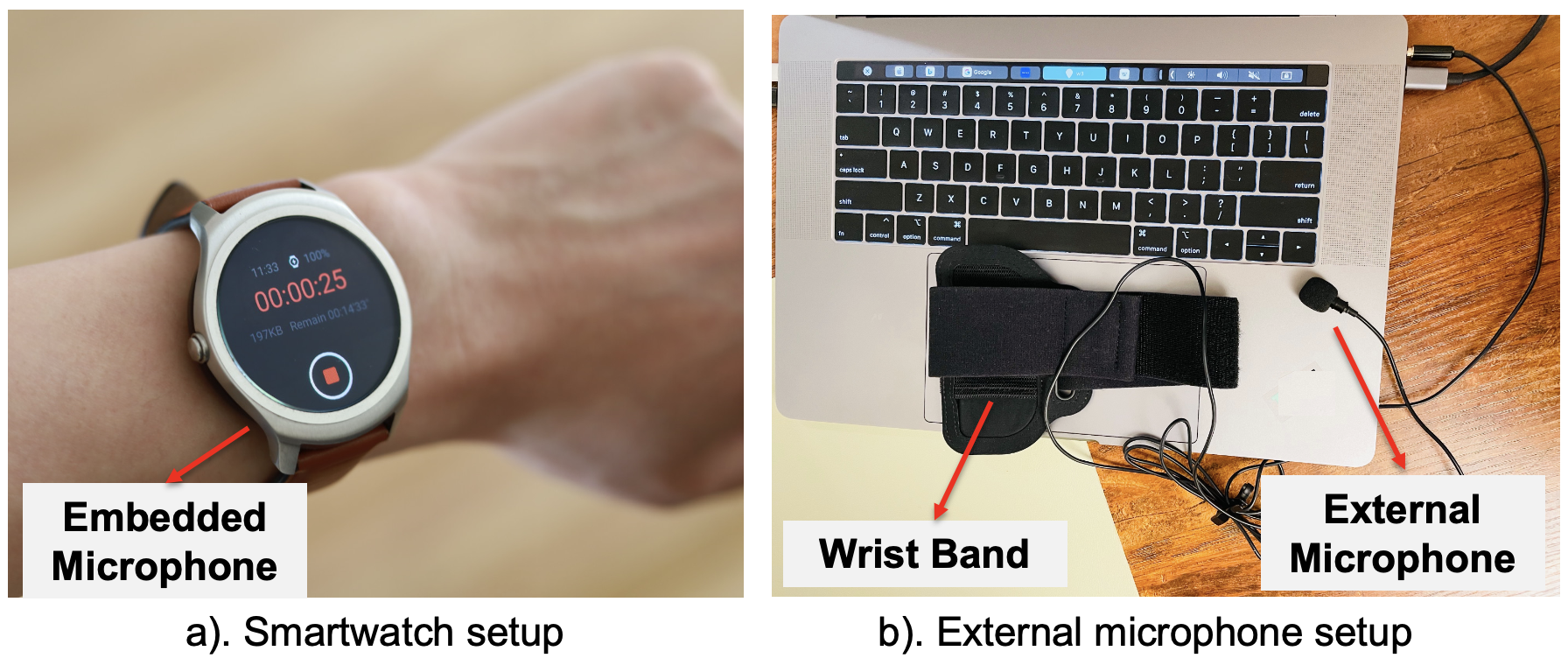}
    \caption{Experiment setup for sound signal capturing and gesture detection. We record signals from both a smartwatch and an external microphone, and run the detection pipeline on a personal computer for evaluation purposes.}
    \label{fig:setup}
\end{figure}

\textbf{Experiment Setup.} To validate the design, we develop a pipeline to capture audio data from an Android smartwatch or an external microphone  attached to the wrist, and run the gesture detection pipeline on a personal computer (shown in Figure \ref{fig:setup}). For the convenience of evaluating the accuracy performance, the recognition pipeline is developed using Python and tested on a personal computer. We also develop two different applications (a web browser and a 3D object manipulation app) to demonstrate how to interact with different user experiences via hand gestures.
%The prototype consists of three major modules: acoustic signal capturing, feature engineering and machine learning pipeline for gesture detection.

\textbf{Machine Learning Pipeline.} The machine learning pipeline requires offline training and online recognition. We record sound signals from a smartwatch using the built-in audio recorder and export the recordings for analysis. For the external microphone, we use pyaudio library for sound recording for training as well as streaming real-time sound signals for real-time gesture detection. We train the CNN model off-line on a computing server equipped with GPUs. Keras~\cite{chollet2015keras} with Tensorflow~\cite{abadi2016tensorflow} backend is used for CNN construction and training. Adam optimizer~\cite{kingma2014adam} with a learning rate of $l=0.001$ is used to speed up the training. For real-time recognition, we run the model inference on a regular personal computer as part of two different applications to test the accuracy of our gesture recognition.

% \subsection{Gesture Detection Modes}
% We propose two detection modes to balance the power consumption: two-factor one-pass authentication, low-power continuous authentication and ultra low-power presence detection, suitable for scenarios requiring progressively less security level but more user convenience and power efficiency.

% % \begin{figure}
% %     \centering
% %     \includegraphics[width=3in]{figures/mode.png}
% %     \caption{Two pass detection mode.}
% %     \label{fig:mode}
% % \end{figure}

% \textbf{One-pass Detection.}
%  This incurs the most computation, energy costs, providing the highest security level suitable for scenarios such as phone unlock, account log in. 

% \textbf{Low-power Two-pass Detection.}
% In this mode, acoustic features extracted from the CNN are used in one-class SVM classification. It provides reduced accuracy level suitable for scenarios such as continuous access/browse of.

% \textbf{``Always On'' low-power Detection}.
% This mode uses acoustic signal only and an SVM model to detect the presence of the user face. To minimize computation and energy costs, the spectrum of a set of samples (e.g., the first 80 after the direct path signal) instead of CNN extracted features is fed to SVM.
\section{Evaluation}
\subsection{Data Collection}
To generate the training data, we invited 5 participants, of different ages and genders, to contribute audio clips using either an Android smartwatch or an external microphone attached to the wrist (shown in Figure \ref{fig:setup}).
The differences in the sound capture equipment of various participants help create a sufficiently diverse and rich data set to create a robust model\footnote{Due to the pandemic, we had to limit the scale of our experiments.}. In addition, the data captured from smartwatches demonstrate the feasibility of our system leveraging sound data from actual wearable devices, which can be used to interact with not only the wearable device itself but also perform 2D and 3D interaction in augmented and virtual reality applications.

During data collection, we instruct the participants to perform each gesture repeatedly. For convenient segmentation, the data is collected in relatively quiet environments. In total, the data set contains 2683 valid samples from 5 gesture classes after the automatic data segmentation and labelling. We randomly shuffle the data and divide it in three parts, $70\%$ for model training, $10\%$ for model validation and $20\%$ testing. After doing the division, we augment each sample with random background noise and ensure that each original sample belongs to just one group (train, test or validation). We test a different number of synthetic samples from each gesture sample, as explained in \ref{sec:eva}, and end up choosing a ratio of 1:10, which yields the best result. Thus we generate 26830 valid samples for 5 gestures in total. 

\begin{table}[h]
\centering
\caption{Evaluation data set summary.}
\begin{tabular}{|c|c|c|c|c|}
  \hline
   & Train & Validation & Test  & Total \\
   \hline
   	Clean  &  1878    & 268     & 537 & 2683 \\
   	 \hline
   	Augmented   & 18780      & 2680     & 5370 & 26830  \\
   	 \hline
\end{tabular}\label{tab_data}
\end{table}

\subsection{System Evaluation}
We train two models based on the clean training set and the augmented training set, and compare the performance on both clean testing set and augmented testing set (which have not been used during training, neither as training samples nor to tune the hyper-parameters of the model). Note that the augmented set is more challenging as it contains background noise, and thus, is a better metric of performance in a real environment deployment.

%We train our model on the clean training data set without augmentation. Then we evaluate the same model on the augmented testing data set, which has synthetic clips created with noise in the background, which makes it a more difficult data set. 
\subsubsection{Recognition Accuracy Evaluation}
\textbf{Overall Accuracy.} We introduce precision, recall, F1-score and accuracy as metrics. Precision is the fraction of true positives among all samples classified as positive, defined as $P = \frac{TP}{TP + FP}$; recall is the fraction of true positives among all positive samples, defined as $R  = \frac{TP}{TP + FN}$. A high precision means the actual gestures can be detected correctly, a high recall means less likely the actual gestures are missed. We also introduce F1-score, which is the harmonic mean of precision and recall defined as $\text{F1} = 2  \frac{P \cdot R}{P + R}$. Accuracy is the proportion of correct predictions (both true positives and true negatives) among the total number of cases examined, defined as $ACC = \frac{TP+TN}{TP+TN+FP+FN}$.

\begin{table}[h]
\centering
\caption{Results summary over test set.}
\begin{tabular}{|c|c|c|c|c|c|}
  \hline
   Train & Test & Precision & Recall & F1  & Accuracy \\
   \hline
   	Clean  & Clean  &  0.9155    & \textbf{0.8883}   & 0.9017 & 0.9013 \\
   	\hline
   	Clean  & Aug.  &  0.4783    & 0.4453     & 0.4612 & 0.4648 \\
   	 \hline
   	Aug. & Clean  & \textbf{0.9293}      & 0.8808     & \textbf{0.9044} & \textbf{0.9088}  \\
   	\hline
   	Aug. & Aug.  & 0.8874      & 0.8320     & 0.8588 & 0.8579  \\
   	 \hline
\end{tabular}
\label{table:results_augmented}
\end{table}

Table \ref{table:results_augmented} shows the results with different combinations, averaging over all classes. In the ideal case, when we train and evaluate the model on the clean data set, the model achieves very promising results of 91.55\% precision and 88.83\% recall, with an F1-score of over 90\%. However, the performance drops significantly when testing on the augmented testing data with background noise. The F1-score drops to 46\%, which is not usable in a real environment. This is mainly caused by the model being trained on samples with minimum background noise, thus not generalizing well on real samples in noisy environments. In contrast, the model trained with augmented training data shows comparable results with the model trained on clean data, when tested against clean test data, while the results of the F1 metric measured on the augmented test data remain above 85\%, a huge improvement compared to the results without augmentation.

\begin{figure}
	\centering
	\includegraphics[width=2.4in]{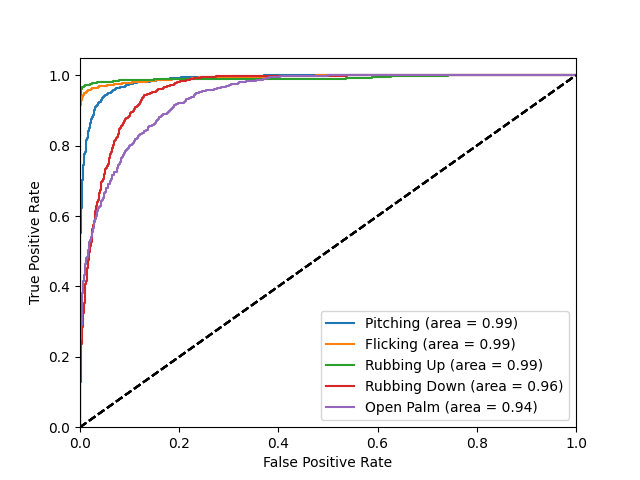}
	\caption{ROC curves for each gesture.}\label{fig_roc}
\end{figure}

\textbf{Per Gesture Accuracy.} 
Different hand gestures involve different finger motions; some are more distinct to recognize and some are less distinct. Thus, we also evaluate the performance for each type of gesture. We show the receiver operating characteristic (ROC) curve and the area under the curve (AUC) for each gesture in our evaluation in Figure \ref{fig_roc}. A ROC curve is a plot that illustrates the diagnostic ability of a classifier system as its threshold is varied. An AUC indicates better classification performance. From Figure \ref{fig_roc}, we can see the pitching, flicking, and rubbing up gestures have relatively larger AUC, which means they are easier to recognize; while rubbing down and open palm gestures have relatively lower performance. The open palm gesture has the worst accuracy among all the gestures, which could be caused by the complex nature of the gesture. It involves two sub-sequential gestures: make a fist and then open the palm. Training a recurrent model could better take advantage of such sequential data for improvement. %Figure \ref{table:results_perclass} 

%The performance has some variation between the different gestures to recognize, from our results it's clear that the gestures 'up' and 'back' works much better than the gesture for 'home'. Table~\ref{table:results_perclass} and figure~\ref{fig_roc} show this variation. These results intuitively make sense, the gesture for 'home' it's a more complex gesture which require the movement of several fingers, which can be interpreted as the concatenation of several simpler gestures which requires the movement of separate fingers.

% \begin{table}[h]
% \centering
% \caption{Performance per class. Using model trained with augmented data and 10 samples. Tested over augmented data.}
% \begin{tabular}{|c|c|c|c|c|}
%   \hline
%   Class  & Precision & Recall & F1  & Support \\
%   \hline
%   	Pitching   &  0.91    & 0.88   & 0.90 & 1170 \\
%   	\hline
%   	Flicking    &  0.97    & 0.94     & 0.95 & 1170 \\
%   	 \hline
%   	Rubbing Up  & \textbf{0.98}      & \textbf{0.96}     & \textbf{0.97} & 1020  \\
%   	\hline
%   	Rubbing Down & 0.67      & 0.90     & 0.77 & 1090  \\
%   	 \hline
%   	 Open Palm & 0.82      & 0.55     & 0.66 & 920  \\
%   	 \hline
% \end{tabular}
% \label{table:results_perclass}
% \end{table}

%To improve the robustness of the model against background noise, we also train the model on the augmented training dataset. We evaluate the improvement in the data augmentation evaluation section.

% \begin{figure}
%     \centering
%     \includegraphics[width=2in]{confuse.png}
%     \caption{Confusion matrix of clean data.}
%     \label{fig:clean_noise}
% \end{figure}

\subsubsection{Data Augmentation Evaluation}\label{sec:eva}
We evaluate how effective data augmentation can improve the performance by generating ``synthesized'' training samples when training data is limited. 
%We augment the training and validation samples, combining the activation clips with background noise. For each activation clip we added a random segment of a background clip, and 
We evaluate the performance regarding to different augmentation ratio, i.e., the ratio between the synthesized samples and the clean samples. Varying this ratio gives different amounts of training samples. Figure \ref{number_samples} shows the results of the model trained with different augmentation ratios, ranging from 1 to 100, and tested against clean and augmented testing data set. As the ratio increases, we observe the performance increases as well. The model trained with the ratio of 10 performs best on the clean test data with a F1-score above 90\%. Although the accuracy is slightly lower than model trained with a ratio of 50 evaluated on the augmented test data, the difference is negligible. However, augmenting the training data with a ratio of 50 significantly increased the data set size, which takes more resources for training. Choosing a even larger ratio of 100, the performance starts to drop as the model could be over-fitted to the background noise rather than the limited gesture sound samples. In our final design, we choose a ratio of 10 to make the trade offs among these factors. 
%We tested creating different number of augmented samples as we had multiple background clips and each one of them was much longer that our activation's clips, making possible to create multiple sets of the new combining clips, in 
%Table \ref{number_samples} we show our results using a different number of augmentations as part of our training data and ended picking a ratio of 1:10, which gave the best results. The data set is shuffled to make it more balanced. We compare the model trained with data augmentation and without augmentation by evaluating the performance on the augmented testing set. The results are shown in Table \ref{table:results_augmented}, averaging over all classes. The model trained on the augmented dataset performs the best, although the difference is negligible when compared with the model trained and tested on clean data. The difference is much more important when there is background noise in the testing dataset, which mimics the real life situation where our model will be used.

\begin{table}[h]
\centering
\caption{Results with different data augmentation ratios.}
\begin{tabular}{|c|c|c|c|c|c|}
  \hline
  Ratio & Test & Precision  & Recall & F1 & Accuracy \\
   \hline
   	1  & Clean     & 0.8167     & 0.7300 & 0.7709 & 0.7728\\
   	 \hline
   	2     & Clean      & 0.8435     & 0.7728 & 0.8066 & 0.8063   \\
   	 \hline
	5    & Clean     & 0.9291     & 0.8547 & 0.8903 & 0.8920 \\
	\hline
	10    & Clean       & \textbf{0.9293}     & \textbf{0.8803} & \textbf{0.9044} & \textbf{0.9088} \\
  	\hline
  	50    & Clean     & 0.8953     & 0.8436 & 0.8687 & 0.8622 \\
  	\hline
  	100    & Clean     & 0.9213     & 0.8715 & 0.8957 & 0.8827 \\
	\hline
   	1  & Aug.     & 0.8308     & 0.7553 & 0.7913 & 0.7935\\
   	 \hline
   	2     & Aug.      & 0.8467     & 0.7952 & 0.8201 & 0.8194   \\
   	 \hline
	5    & Aug.     & 0.8601     & 0.8233 & 0.8413 & 0.8389 \\
	\hline
	10    & Aug.       & 0.8874     & 0.8320 & 0.8588 & 0.8579 \\
  	\hline
	50    & Aug.     & \textbf{0.8909}     & \textbf{0.8456} & \textbf{0.8677} & \textbf{0.8654} \\
	\hline
	100    & Aug.     & 0.8795     & 0.8292 & 0.8536 & 0.851 \\
	\hline
	
\end{tabular}\label{number_samples}
\end{table}

%We also vary the augmentation ratio to generate different amount of augmented training data. As we can see, the performance increases as we have more and more data and reach a plateau when the ratio reaches xx.
% It is obvious that data augmentation improves recall significantly, thus F-score and BAC, especially when the training samples are very limited (e.g., $<$100). As the size grows, the recall with data augmentation is always higher. However the precision decreases to $\sim95\%$, which is because ``synthesized'' training samples have more noises, making it easier to have false positives. The performance becomes stable with more than 400 training samples, which can be collected within one minute when registering a new user.

% \begin{figure}
% 	\centering
% 	\subfigure[Without DA.] {
% 		\includegraphics[width=1.58in]{eva_wo_da.pdf}\label{fig_wo_da}
% 		}
% 	\subfigure[With DA.]{
% 		\includegraphics[width=1.58in]{eva_w_da.pdf}\label{fig_w_da}
% 	}
% 	\caption{Classification performance comparison of data augmentation (DA) under different training data amounts.} \label{fig_eva_da}
% \end{figure}

\subsubsection{False Alarm Rate Evaluation}
\begin{figure}
	\centering
	\includegraphics[width=2.4in]{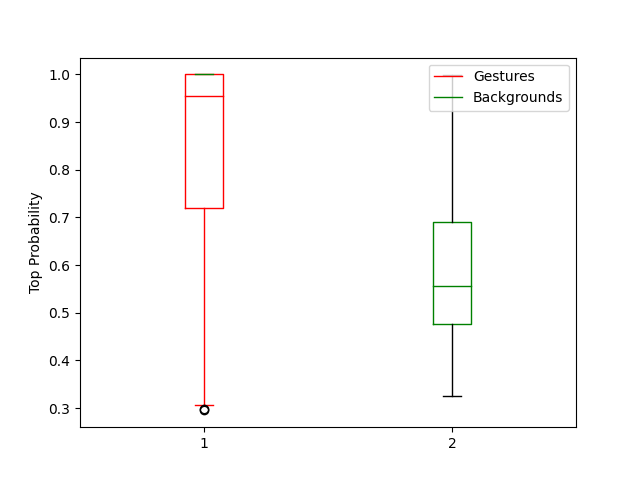}
	\caption{Distributions of probabilities of events triggered by finger gestures and background noises.}\label{fig_boxplot}
\end{figure}

Choosing the threshold $\epsilon$ is critical to make the balance between true positives and false alarms. It is important to keep the false alarm rate as low as possible to avoid unintended input signals. A false alarm is any of the 5 gestures being detected while the user does not intend to make it. Figure~\ref{fig_boxplot} shows the probability distribution of the events triggered by actual finger gestures and background noise. As we can see, the probabilities triggered by actual gestures are usually much higher than those triggered by background noise. Choosing a threshold $\epsilon=0.7$ allows us to capture the majority of actual gestures and ignore the background noise events. Thus, we set our primary threshold as 0.7. We choose 0.6 as our secondary threshold to increase the sensitivity in a more challenging environment, as proposed in our dynamic thresholding detection design.
%range of activation's caused by real activation's and the ones caused by plain background noise, based on this we set a high activation threshold (of 0.7) which discarded most of the background noise caused activation's.

\begin{figure}[h]
    \centering
    \includegraphics[width=2.8in]{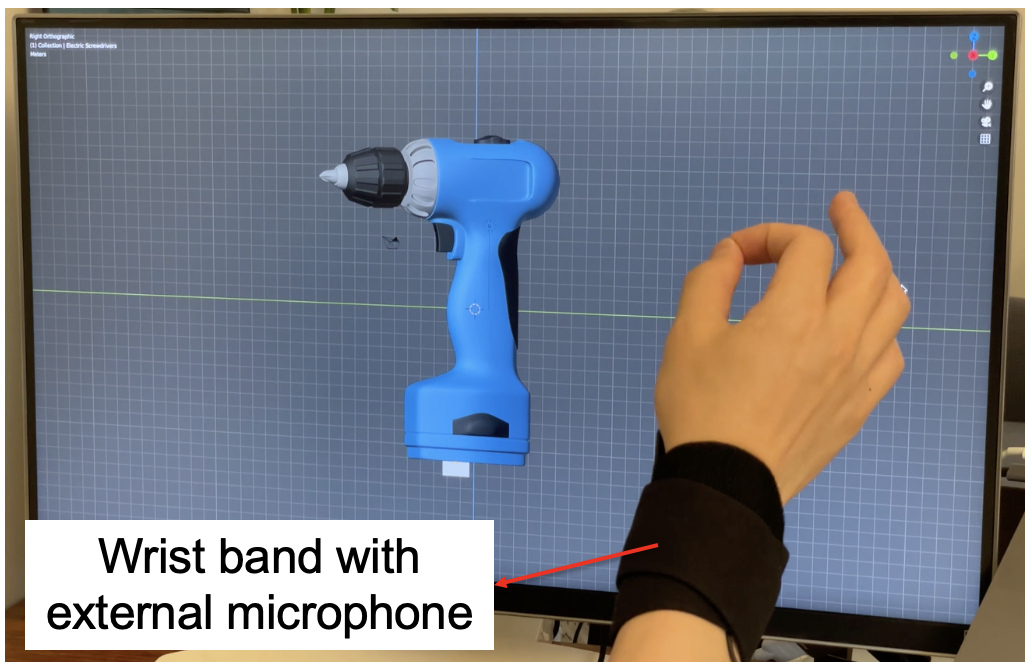}
    \caption{Using finger gestures to manipulate 3D models.}
    \label{fig:application}
\end{figure}

\subsubsection{Qualitative User Experience Study}\label{sec:attack_eval}
 We built a web browser navigation application and a 3D object manipulation application to test our gesture detection system. The mapping of gestures to controls are listed in Table \ref{tab:mapping}.
%\textcolor{red}{PLEASE FILL THE IFNO For the web application, the mapping of the gestures to controls is as described in Section X. For the 3D object manipulation application, we used the following mapping: click -> select object, flick->, .... }
We conduct a survey with 5 users to collect their feedback, mainly on the usability of the system. 

Both applications run in real-time on a personal computer, so that the users can interact with the applications via the hand gestures. Out of the 5 users, 4 reported being able to control the web browser easily and agreed that our gesture-based interaction is useful for such browsing tasks. All of them rated it equally easy to scroll the web pages as when using a mouse. As we are able to successfully capture data from wearable devices to classify hand gestures in real time, all participants agree this system would be very useful for 3D object manipulation in AR/VR settings to eliminate the need for additional controllers. No obvious latency is observed for each gesture input. Figure \ref{fig:application} shows the application in action to control 3D objects on the screen via gestures. By attaching an external microphone to their wrist, the users can easily manipulate a 3D object such as zooming in/out, performing rotation and returning to default view. Occasional mistaken gestures or false positives are observed in noisy environment, which needs improvement. Nevertheless, the participants agree our system provides a new way for interaction with a great potential for broader applications.

\begin{table}
\centering
\caption{Mapping of gestures to controls.}
\begin{tabular}{|c|c|c|}
    \hline
        & Web Browser & 3D Object Viewer\\
    \hline
    Pitching & Reserved & Zoom In \\
    \hline
    Rubbing Up & Scroll Up & Rotate Right \\
    \hline
    Rubbing Down & Scroll Down & Rotate Left \\
    \hline
    Flicking & Go Back & Zoom Out \\
    \hline
    Open Palm & Exit Browser & Default View \\
    \hline
    
%   \hline
%   & Pitching & Scroll Up & Scroll Down & Flicking & Open Palm \\
%   \hline
%   	Web & Clean  & Clean  &  0.9155    & \textbf{0.8883}   & 0.9017  \\
%   	\hline
%   	3D & Zoom In  & Rotate Right  &  Rotate Left    & Zoom Out     & Default View \\
%   	 \hline
\end{tabular}
\label{tab:mapping}
\end{table}

\section{Discussion}
\textbf{Limitations.}
This is only a research prototype and far from a well engineered product. It has several limitations: 
\emph{i) Implementation on wearable devices.} Although we validated the feasibility and performance on a personal computer, and also proposed solutions to meet the computation power and battery constraints on wearable devices, we do not have the implementation of the system on wearable devices yet, other than the sound capturing component for data generation. 
\emph{ii) False alarms.} Although the acoustic data augmentation greatly improved the performance in noisy environments, there is still room for further improvement, especially in very noisy environments. 
\emph{iii) User gesture changes.} The current CNN model is trained on limited data, far from exhaustive to be robust against various hand gesture variations. An online model updating mechanism is needed to address such changes dynamically.
% \emph{iv) continuous detection usability.} Although we demonstrate it has the potential for pure acoustic-based continuous authentication, it requires the smartphone aligned and in front of the user's face, which impacts the usability. To mitigate such problem, we can either decrease the frequency of acoustic sampling and detection at a cost of decreased security, or register users with data from different angles thus enlarging the range of effective detection.

\textbf{Future Work.}
\emph{i) Collecting a larger data set.} We will collect more training data from more users with a larger variety of devices. This will further improve the performance along with a more sophisticated neural network design.
\emph{ii) Wearable device implementation.} We will implement applications on wearable devices and perform more evaluations regarding the computation resources and power consumption.
\emph{iii) Large scale experiment.} Our current data collection and experiments are limited to only $\sim 5$ users. Large scale experiments are needed to improve the maturity of the solution.

\section{Conclusion}
In this paper, we propose a system that leverages bone-conducted sound signals for 2D and 3D interaction. We designed a complete pipeline for robust hand gesture detection even in noisy environments, and validated its usability through different applications. We showed that we are able to successfully capture data from wearable devices to classify hand gestures in real time, which can be used to interact with not only the wearable device itself, but also potentially perform 2D and 3D interaction in augmented and virtual reality applications without the need for controllers. Experiments show that our system achieves an overall accuracy of 90.13\% in a quiet environment, and 85.79\% under noisy conditions. 

% \newpage
% \noindent\fbox{%
%     \parbox{\linewidth}{
%     \color{red}
%     \textbf{Important Dates:}
%         \begin{itemize}
%             \item 4/16: Submit disclosure.
%             \item 5/01: Get disclosure evaluated and clearance approval.
%             \item 5/24: Conference Papers abstracts due (REQUIRED).
%             \item 5/28: Conference Papers submissions due.
%             \item 7/23: Conference Papers notifications.
%         \end{itemize}
        
%     \textbf{Tentative Milestones:}
%         \begin{itemize}
%             \item 4/23: Finalize the architecture design.
%             \item 4/30: Finish first draft writing and evaluation plan.
%             \item 5/10: Get majority of the evaluation results.
%             \item 5/20: Complete all the results needed and get the paper into 90\%.
%             \item 5/24: Submit abstract.
%             \item 5/28: Polish draft and submit.
%         \end{itemize}
        
%         Page: 4–8 pages of text,  NOT including references. References should not be exceeding 2 pages.
%     }
% }

%\acknowledgments{}

% \clearpage
{\small
\bibliography{bib}

\begin{thebibliography}{26}
\providecommand{\natexlab}[1]{#1}
\providecommand{\url}[1]{\texttt{#1}}
\providecommand{\urlprefix}{URL }
\expandafter\ifx\csname urlstyle\endcsname\relax
  \providecommand{\doi}[1]{doi:\discretionary{}{}{}#1}\else
  \providecommand{\doi}{doi:\discretionary{}{}{}\begingroup
  \urlstyle{rm}\Url}\fi

\bibitem[{hol(5/28/2021)}]{hololens}
 5/28/2021.
\newblock Microsoft HoloLens | Mixed Reality Technology for Business.
\newblock https://www.microsoft.com/en-us/hololens.

\bibitem[{sir(5/28/2021)}]{siri}
 5/28/2021.
\newblock Siri does more than ever. Even before you ask.
\newblock https://www.apple.com/siri/.

\bibitem[{Abadi et~al.(2016)Abadi, Barham, Chen, Chen, Davis, Dean, Devin,
  Ghemawat, Irving, Isard et~al.}]{abadi2016tensorflow}
Abadi, M.; Barham, P.; Chen, J.; Chen, Z.; Davis, A.; Dean, J.; Devin, M.;
  Ghemawat, S.; Irving, G.; Isard, M.; et~al. 2016.
\newblock TensorFlow: A System for Large-Scale Machine Learning.
\newblock In \emph{OSDI}, volume~16, 265--283.

\bibitem[{Chollet et~al.(2015)}]{chollet2015keras}
Chollet, F.; et~al. 2015.
\newblock Keras.
\newblock url{https://github.com/keras-team/keras}.

\bibitem[{Dementyev and Paradiso(2014)}]{dementyev2014wristflex}
Dementyev, A.; and Paradiso, J.~A. 2014.
\newblock WristFlex: low-power gesture input with wrist-worn pressure sensors.
\newblock In \emph{Proceedings of the 27th annual ACM symposium on User
  interface software and technology}, 161--166.

\bibitem[{He et~al.(2016)He, Zhang, Ren, and Sun}]{he2016deep}
He, K.; Zhang, X.; Ren, S.; and Sun, J. 2016.
\newblock Deep residual learning for image recognition.
\newblock In \emph{Proceedings of the IEEE conference on computer vision and
  pattern recognition}, 770--778.

\bibitem[{Hussin, Birasamy, and Hamid(2016)}]{hussin2016design}
Hussin, S.~F.; Birasamy, G.; and Hamid, Z. 2016.
\newblock Design of Butterworth Band-Pass Filter.
\newblock \emph{Politeknik \& Kolej Komuniti Journal of Engineering and
  Technology} 1(1).

\bibitem[{Kingma and Ba(2014)}]{kingma2014adam}
Kingma, D.~P.; and Ba, J. 2014.
\newblock Adam: A method for stochastic optimization.
\newblock \emph{arXiv preprint arXiv:1412.6980} .

\bibitem[{Logan et~al.(2000)}]{logan2000mel}
Logan, B.; et~al. 2000.
\newblock Mel Frequency Cepstral Coefficients for Music Modeling.
\newblock In \emph{ISMIR}, volume 270, 1--11.

\bibitem[{Mueller et~al.(2018)Mueller, Bernard, Sotnychenko, Mehta, Sridhar,
  Casas, and Theobalt}]{mueller2018ganerated}
Mueller, F.; Bernard, F.; Sotnychenko, O.; Mehta, D.; Sridhar, S.; Casas, D.;
  and Theobalt, C. 2018.
\newblock Ganerated hands for real-time 3d hand tracking from monocular rgb.
\newblock In \emph{Proceedings of the IEEE Conference on Computer Vision and
  Pattern Recognition}, 49--59.

\bibitem[{Nandakumar et~al.(2016)Nandakumar, Iyer, Tan, and
  Gollakota}]{nandakumar2016fingerio}
Nandakumar, R.; Iyer, V.; Tan, D.; and Gollakota, S. 2016.
\newblock FingerIO: Using Active Sonar for Fine-Grained Finger Tracking.
\newblock In \emph{Proceedings of the 2016 CHI Conference on Human Factors in
  Computing Systems}, 1515--1525. ACM.

\bibitem[{Neubeck and Van~Gool(2006)}]{neubeck2006efficient}
Neubeck, A.; and Van~Gool, L. 2006.
\newblock Efficient non-maximum suppression.
\newblock In \emph{18th International Conference on Pattern Recognition
  (ICPR'06)}, volume~3, 850--855. IEEE.

\bibitem[{Nguyen et~al.(2019)Nguyen, Rupavatharam, Liu, Howard, and
  Gruteser}]{nguyen2019handsense}
Nguyen, V.; Rupavatharam, S.; Liu, L.; Howard, R.; and Gruteser, M. 2019.
\newblock HandSense: capacitive coupling-based dynamic, micro finger gesture
  recognition.
\newblock In \emph{Proceedings of the 17th Conference on Embedded Networked
  Sensor Systems}, 285--297.

\bibitem[{Ogata and Imai(2015)}]{ogata2015skinwatch}
Ogata, M.; and Imai, M. 2015.
\newblock SkinWatch: skin gesture interaction for smart watch.
\newblock In \emph{Proceedings of the 6th Augmented Human International
  Conference}, 21--24.

\bibitem[{Peng et~al.(2007)Peng, Shen, Zhang, Li, and Tan}]{peng2007beepbeep}
Peng, C.; Shen, G.; Zhang, Y.; Li, Y.; and Tan, K. 2007.
\newblock BeepBeep: A High Accuracy Acoustic Ranging System using COTS Mobile
  Devices.
\newblock In \emph{ACM SenSys}.

\bibitem[{Saponas et~al.(2010)Saponas, Tan, Morris, Turner, and
  Landay}]{saponas2010making}
Saponas, T.~S.; Tan, D.~S.; Morris, D.; Turner, J.; and Landay, J.~A. 2010.
\newblock Making muscle-computer interfaces more practical.
\newblock In \emph{Proceedings of the SIGCHI Conference on Human Factors in
  Computing Systems}, 851--854.

\bibitem[{Shen, Wang, and Roy~Choudhury(2016)}]{shen2016smartwatch}
Shen, S.; Wang, H.; and Roy~Choudhury, R. 2016.
\newblock I am a smartwatch and i can track my user's arm.
\newblock In \emph{Proceedings of the 14th annual international conference on
  Mobile systems, applications, and services}, 85--96.

\bibitem[{Shorten and Khoshgoftaar(2019)}]{shorten2019survey}
Shorten, C.; and Khoshgoftaar, T.~M. 2019.
\newblock A survey on image data augmentation for deep learning.
\newblock \emph{Journal of Big Data} 6(1): 1--48.

\bibitem[{Simonyan and Zisserman(2015)}]{simonyan2014very}
Simonyan, K.; and Zisserman, A. 2015.
\newblock Very Deep Convolutional Networks for Large-Scale Image Recognition.
\newblock In \emph{International Conference on Learning Representations}.

\bibitem[{Sridhar, Oulasvirta, and Theobalt(2013)}]{sridhar2013interactive}
Sridhar, S.; Oulasvirta, A.; and Theobalt, C. 2013.
\newblock Interactive markerless articulated hand motion tracking using RGB and
  depth data.
\newblock In \emph{Proceedings of the IEEE international conference on computer
  vision}, 2456--2463.

\bibitem[{Wang, Liu, and Sun(2016)}]{wang2016device}
Wang, W.; Liu, A.~X.; and Sun, K. 2016.
\newblock Device-free gesture tracking using acoustic signals.
\newblock In \emph{Proceedings of the 22nd Annual International Conference on
  Mobile Computing and Networking}, 82--94. ACM.

\bibitem[{Weichert et~al.(2013)Weichert, Bachmann, Rudak, and
  Fisseler}]{weichert2013analysis}
Weichert, F.; Bachmann, D.; Rudak, B.; and Fisseler, D. 2013.
\newblock Analysis of the accuracy and robustness of the leap motion
  controller.
\newblock \emph{Sensors} 13(5): 6380--6393.

\bibitem[{Wen, Ramos~Rojas, and Dey(2016)}]{wen2016serendipity}
Wen, H.; Ramos~Rojas, J.; and Dey, A.~K. 2016.
\newblock Serendipity: Finger gesture recognition using an off-the-shelf
  smartwatch.
\newblock In \emph{Proceedings of the 2016 CHI Conference on Human Factors in
  Computing Systems}, 3847--3851.

\bibitem[{Wilhelm et~al.(2015)Wilhelm, Krakowczyk, Trollmann, and
  Albayrak}]{wilhelm2015ering}
Wilhelm, M.; Krakowczyk, D.; Trollmann, F.; and Albayrak, S. 2015.
\newblock eRing: multiple finger gesture recognition with one ring using an
  electric field.
\newblock In \emph{Proceedings of the 2nd international Workshop on
  Sensor-based Activity Recognition and Interaction}, 1--6.

\bibitem[{zapsplat.com(2021)}]{zapsplat}
zapsplat.com. 2021.
\newblock Royalty free sound effects.
\newblock \url{https://www.zapsplat.com}.
\newblock Accessed: 2021-05-20.

\bibitem[{Zhou et~al.(2017)Zhou, Elbadry, Gao, and Ye}]{zhou2017batmapper}
Zhou, B.; Elbadry, M.; Gao, R.; and Ye, F. 2017.
\newblock BatMapper: Acoustic Sensing Based Indoor Floor Plan Construction
  Using Smartphones.
\newblock In \emph{Proceedings of the 15th Annual International Conference on
  Mobile Systems, Applications, and Services}, 42--55. ACM.

\end{thebibliography}
}
\end{document}